# Comparison of Update and Genetic Training Algorithms in a Memristor Crossbar Perceptron

Kyle N. Edwards and Xiao Shen[*]

Department of Physics and Materials Science, University of Memphis, Memphis, TN 38152, USA

**Abstract**

Memristor-based computer architectures are becoming more attractive as a possible choice of hardware for the implementation of neural networks. However, at present, memristor technologies are susceptible to a variety of failure modes, a serious concern in any application where regular access to the hardware may not be expected or even possible. In this study, we investigate whether certain training algorithms may be more resilient to particular hardware failure modes, and therefore more suitable for use in those applications. We implement two training algorithms—a local update scheme and a genetic algorithm—in a simulated memristor crossbar, and compare their ability to train for a simple image classification task as an increasing number of memristors fail to adjust their conductance. We demonstrate that there is a clear distinction between the two algorithms in several measures of the rate of failure to train.

In recent years, adjustable two-terminal resistive devices, or memristors, have gained increasing attention as a possible alternative to traditional CMOS transistor-based integrated circuits. Such devices are capable of performing the computational operations traditionally executed on transistors with a von Neumann architecture,[1–3] while also allowing the development of novel structures suitable for non-traditional methods of computation. For example, with the "CrossNet" described in Ref. 4, one may readily implement a perceptron network in a memristor crossbar by encoding the input patterns as voltages across the rows of the crossbar and summing the resultant currents along the columns, functionally using Ohm's Law to perform the inner products of the perceptron. The output currents may then be passed through CMOS support circuitry to determine the appropriate update scheme for the conductance values of the memristors for the next iteration of the algorithm.[5] Memristor crossbars have also been used to find solutions to partial differential equations by implementing a variety of numerical methods on the crossbar.[3] Such devices and methods are part of the "neuromorphic" computing paradigm, in which the von Neumann architecture that has been the dominant paradigm since the advent of modern computation is replaced by an architecture intended to mimic the function of biological neural networks, trading raw speed for greater connectivity of the basic elements of computation.[6,7]

---

[*] Email: xshen1@memphis.edu

However, memristors are susceptible to defects in ways different from conventional semiconductor circuits. Individual memristors in a crossbar may become "stuck" in the low-resistance state after the application of many voltage pulse update cycles,[8] and improper doping can also cause failure to switch between on and off states.[9] Here we investigate the effects of memristor failure in a memristor perceptron, a simple neural network described in Refs. 4–6. Though it has been shown that intentional removal of particular nodes can be of benefit to network training,[10] the influence of uncontrolled node failure is less clear. Certainly, a crossbar with all memristors damaged will fail to train. In this study, we are interested in the rate of failure to train in the device, within "reasonable time", with increasing damage. It can be expected that, as the number of damaged memristors increases, the performance of the perceptron will deteriorate rapidly, and we are furthermore interested to know if a different algorithm, implemented on the same device, will exhibit a significantly different failure curve.

There have been many schemes proposed and utilized for training neural networks,[11,12] and many of these have been implemented on crossbars.[5,13] One broad distinction to be made within these schemes is between methods that search the network's configuration space locally, such as gradient descent methods,[14] and those that search globally, such as genetic algorithms. Genetic algorithms are often used in training neutral networks[15-17] and also have been suggested for use in neuromorphic hardware.[18,19] Although these may be more difficult to implement in memristor crossbars, it is generally assumed that genetic algorithms can potentially provide better convergence in situations where other methods may get stuck in local minima. It is not clear if local or global schemes will be more suitable for neuromorphic hardware such as memristor crossbars, in which failure of the nodes may occur. In this study, we begin addressing this question through a comparison of an update method (Ref. 5) as an example of the local schemes and a simple genetic algorithm representing the global schemes.

In order to investigate the question above, both update and genetic training algorithms are implemented in a simulation of a memristor crossbar, shown in Fig. 1, and trained on a simple image classification task similar to that in Ref. 5. With the expectation that a small input pattern (and therefore a small crossbar) would decay too rapidly to provide useful data, the size of the training patterns is increased from 3×3 pixels in Ref. 5 to 6×6 pixels, and two new categories of letter are introduced. As in Ref. 5, the network consists of only one layer. To accommodate the enlarged image and new categories, the network has been expanded compared to the one used in Ref. 5. Given 36 image pixels, plus the "bias pixel" and a total of 5 image categories, a suitable perceptron architecture for this task would consist of 185 weights. This number is doubled (in accordance with Ref. 5) to a total of 370 memristors for our network by the addition of a second column of memristors for each image category so that the conductances of adjacent columns may be treated as differential pairs to compose the network's weights. The training set consists of the ideal patterns depicted in Fig. 2 (a)-(e) and all possible one-pixel inversions, three of which are depicted in Fig. 2 (h)-(j). The test set consists of the ideal patterns and all two-element inversions, two of which are shown in Fig. 2 (f) and (g). The training and test sets do not change with the number of damaged memristors.

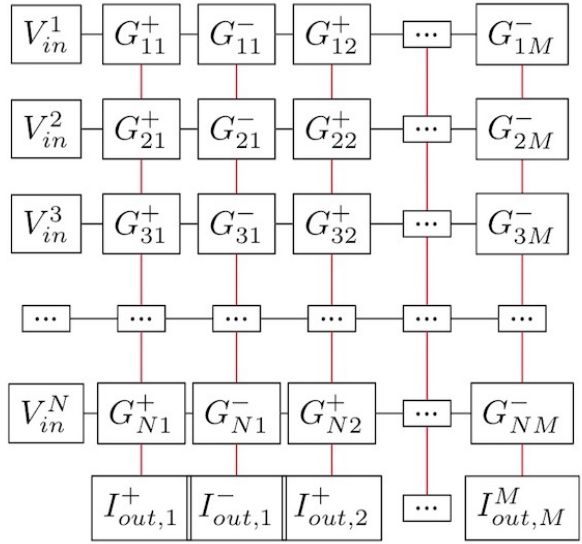

Fig. 1: Simple model of the memristor crossbar. At each node, the horizontal, black wires contact one of the two terminals of the memristor, while the vertical, red wires contact the other terminal.

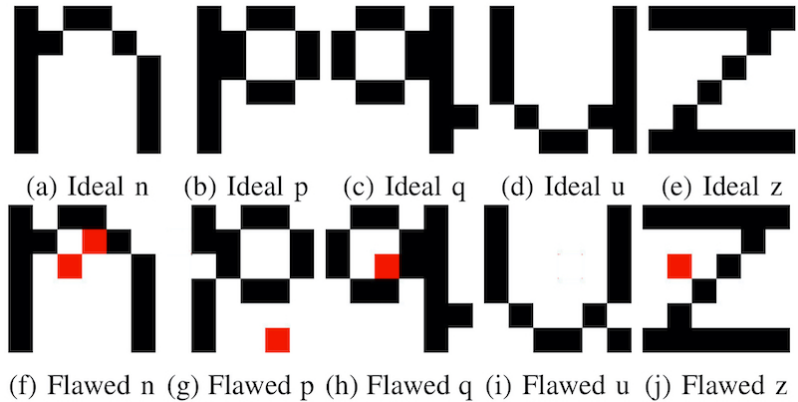

(a) Ideal n  (b) Ideal p  (c) Ideal q  (d) Ideal u  (e) Ideal z

(f) Flawed n  (g) Flawed p  (h) Flawed q  (i) Flawed u  (j) Flawed z

Fig. 2: (a)-(e): The ideal representation of the five letter categories. (f)-(j): An example of a flawed representation for each of the categories. The flawed n and p depict two-cell inversions, from the test data, while the q, u, and z depict one-cell inversions, from the training data.

The update method used closely follows the Manhattan rule used in Ref. 5. Rather than calculating an adjustment to the conductance of each memristor based on the error gradient, only the direction of change is determined for each row of the crossbar based on the perceptron's classification of the test data. Then, an update voltage pulse of fixed magnitude is simulated to be applied to each row, the polarity of which corresponds to the calculated direction. The effect of each update pulse on the conductance of the memristors in that row is determined by a model derived from experimental data (equations S1 and S2 in the supplement of Ref. 5), constrained within the physical limits of 10 μS and 100 μS. The conductance ($G$) increase used in this paper takes the form

$$\Delta G_{set}(G) = 10^{-3} \left(10^6 G - 10^6 G_{min} + 10^{\frac{vset}{2}}\right)^{-2}, \qquad (1)$$

and the decrease has the form

$$\Delta G_{reset}(G) = -10^{-3} \left(-10^6 G + 10^6 G_{max} + 10^{\frac{vreset}{2}}\right)^{-2}, \qquad (2)$$

where *vset* and *vreset* are random values chosen between [1,5.5] to simulate variation in switching threshold and $G_{max}$ and $G_{min}$ represent the physical conductance bounds of the devices.

Broadly, a genetic algorithm mimics biological evolutionary processes by randomly generating a number of possible solutions, called chromosomes, and evaluating their performance against a certain measure of fitness. Those that perform well are subjected to some degree of further randomization, or mutation, and then "bred" together in a process that mixes some amount of the "parent" chromosomes together in offspring chromosomes. This new set of proposed solutions, both parents and offspring, is again evaluated for fitness, and the process is repeated until an acceptable solution is found. A single crossbar can hold only one chromosome at a time, which requires the genetic algorithm's search to be performed sequentially. This raises the question of how to store the non-active chromosomes. Since part of the neural network's processing must be handled by the CMOS support circuitry, the conductance values of that active chromosome can be read by small "read" voltage pulses and stored on that same support circuitry. Then, when necessary, the chromosomes can be switched out by the application of appropriate "write" voltage pulses across the crossbar.[4,20,21]

Our genetic training algorithm first generates random conductance matrices as its chromosomes, corresponding to crossbars with random initial memristor values. The matrices' conductance values are altered or swapped according to the algorithm's rules, which in a physical device would be achieved by voltage pulses applied to the appropriate device, much like the update scheme. The genetic scheme also follows a similar algorithm to the update scheme to produce synaptic outputs. Specifically, given the *n*th conductance matrix $G^n$ with elements $G^n_{jk}$ and training vector $\bar{v}^i$ with elements $v^i_j$, we calculate the *i*th element of the output vector $\bar{u}^i$, with elements $u^i_j$, as follows, corresponding to the product of matrix $G$ with $\bar{v}^i$:

$$u_j^i = \sum_{k=1}^{37} G_{jk}^n v_j^i. \tag{3}$$

Physically, these outputs may be interpreted as the currents resulting from voltage pulses applied to the inputs of the crossbar, determined by the training patterns. Each output vector $\bar{u}^i$ is then passed, element-wise, through an activation function, producing output vector $\bar{f}^i$ with elements:

$$f_j^i = \tan(\beta u_j^i), \tag{4}$$

where $\beta$ is a scaling factor of $2\times10^5\,\text{A}^{-1}$ to center the argument around 1. These synaptic outputs for each training pattern, calculated identically to those in the update method, are then used to find the cost of each chromosome, namely the mean-square of the error of each training vector's response from the ideal output for that letter category:

$$\text{Cost}(G^n) = \sqrt{\frac{1}{185}\sum_{i=1}^{185} E(i)}, \tag{5}$$

where the error of the $i$th training pattern's output from the ideal, $E(i)$, is

$$E(i) = \sum_{j=1}^{5} D^2(j), \tag{6}$$

and the element-wise difference $D(j)$, between an output vector corresponding to the $m$th category of letter and the ideal is

$$D(j) = \begin{cases} 2 - f_j^i & \text{if } m \text{ divides } i \\ f_j^i & \text{otherwiese} \end{cases}. \tag{7}$$

The magnitude of the ideal output vector is chosen simply to be large relative to the size of a typical output value.

With the baseline behavior established, damage to the synapses is simulated by fixing the value of a given number of randomly selected memristors at a randomly chosen, physically realistic value of conductance. For each number of damaged memristors, 100 trials are run, during which each algorithm is allowed to generate or update an equal number of matrices. Any trials during which the perceptron does not manage to achieve the perfect classification of the training set after producing the allowed number of matrices is considered a failure. The total number of failed trials and the average number of matrices produced before perfect classification is achieved are recorded for each number of damaged memristors, and the results are shown in Fig. 3.

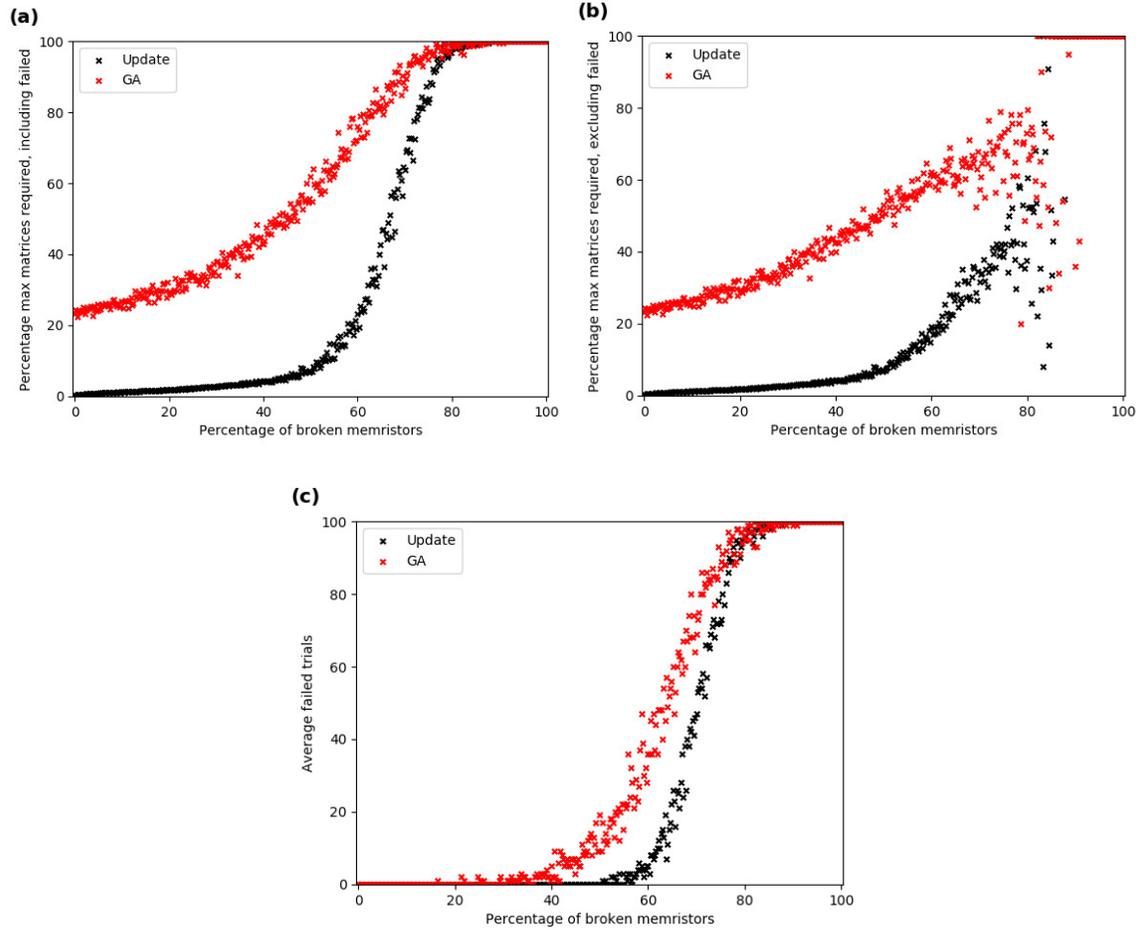

Fig. 3: Results of training the update and genetic algorithms (GA) averaged across one hundred trials, for a variable number of broken memristors. (a) Average fraction of matrices required to train, including failed trials. (b) Average fraction of matrices required to train, excluding failed trials. (c) Average number of failed trials. Note the behavior in the tail of Fig. 3b is due to the steady reduction in number of trials over which the average is taken. By the end, most of these runs are "averaged" over a single trial. Those for which all trials failed to train are defined to have required 100 trials.

Both algorithms are successful for a few broken memristors, as expected from Ref. 10. However, as the damage grows more extensively, the average number of matrices generated before training is achieved steadily increases, as seen in Fig. 3a and Fig. 3b, and a few trials begin to fail entirely. For the update method, the weight matrices in such trials typically feature several memristors that had capped out at the physically imposed clipping values of 10 μS and 100 μS. As the number of broken memristors increases, those memristors which could still be updated become increasingly saturated with the capped values, until eventually all 100 trials consistently fail for each number of broken memristors, as seen in Fig. 4a.

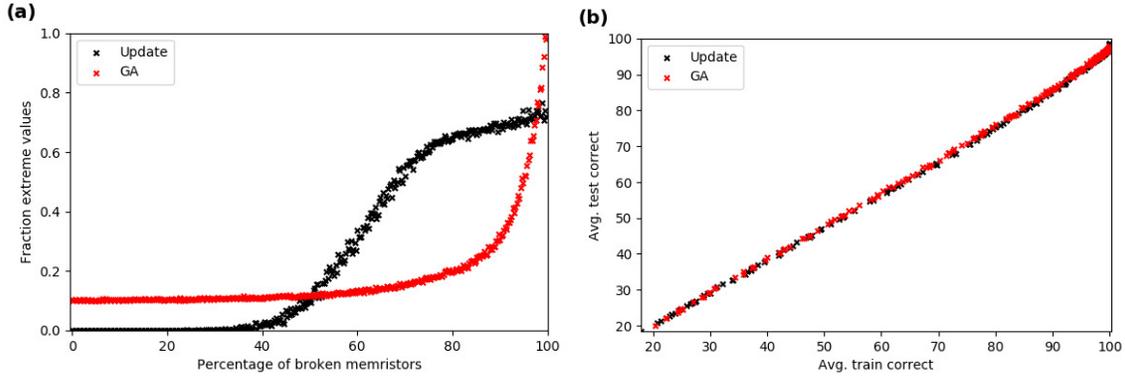

Fig. 4: (a) Cohort average of the fraction of non-damaged memristors which had settled within 5% of the clipping values by the end of training. (b) Comparison of the two algorithms in terms of ability to generalize in spite of damage. Performance is nearly identical, suggesting that the choice of these two particular algorithms has little effect on the ability of a given solution to generalize, except perhaps by inherent resilience to damage.

The genetic algorithm experiences qualitatively similar failure. The minimally-guided, broadly random search for a suitable weight matrix performs well at the beginning of the simulation, but as more conductances are locked in place by damage, the genetic algorithm is no longer able to search the space of possible matrices with the freedom necessary to reliably generate solutions. Similar to the update method, those memristors remaining undamaged settle on values near the extremes of the range of allowed conductances.

The update method demonstrates clear superiority to our simple genetic algorithm in the face of increasing damage. The genetic algorithm begins to suffer a rapid increase in failure about 75 memristors earlier than the update scheme, as shown in Fig. 3c, and the update method maintains more rapid and more reliable training throughout. Though it is interesting to note the similarity between the tail end of the failure curves in Fig. 3, we have shown conclusively that, particularly in the early phase of increasing device failure, there is a definite advantage to using one training algorithm over another in the context of device failure. Given these results, the obvious next question is to determine under which circumstances certain algorithms might respond better to damage than others. Certainly, we might expect that a better-designed genetic algorithm could outperform the update scheme used in this report, but can we find some situation in which a particular class of algorithm is superior to all others for this simple classification task? Perhaps a failure mode other than the one used in this paper could evoke this behavior. Furthermore, what of more advanced machine learning tasks and methods? As discussed in Ref. 4, the CrossNet can be applied to a greater range of machine learning tasks, especially those that are well-suited for a neural net architecture. It may be interesting to explore if different algorithms might exhibit superior resilience under failure for such tasks. Fig. 4b indicates that, at least for this choice of task and algorithms, the ability of the network to generalize from the training data, which is one of the most pertinent metrics of a network's performance, differs little between the algorithms under increasing damage. A potential avenue for future research is further investigation of this aspect of performance, testing if algorithms might be found that do perform better in this regard.

Though this work does not initially aim to investigate the classification performance of the networks, the behavior displayed in Fig. 5 raises an interesting question for future work. In both Fig. 5a and Fig. 5b, the classification performance of the genetic algorithm degrades sooner with increasing failure than the update scheme, in both training and testing. This suggests that it might be interesting to also investigate classification for various training algorithms, and this might be further extended with testing under different failure modes. Another line of future research could be comparing the algorithms under failure modes that a memristor device might experience other than becoming stuck at a particular random value.

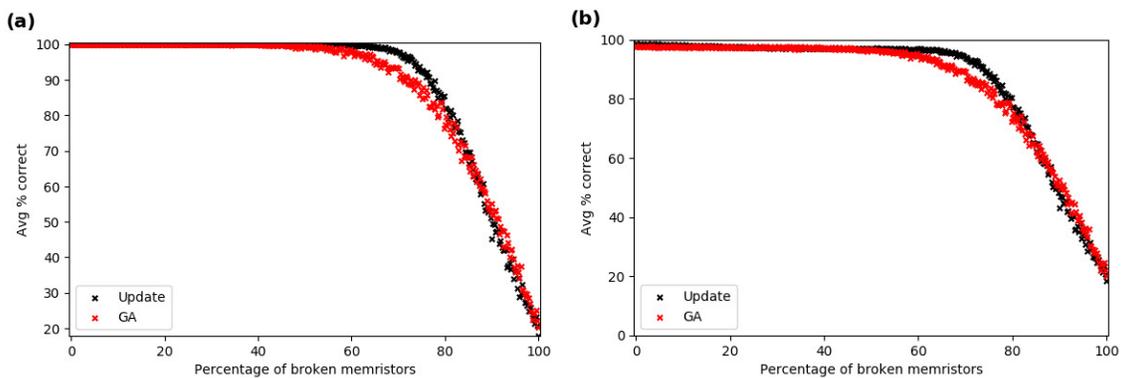

Fig. 5: Classification performance for training (a) and testing (b) data for given number of broken memristors.

It is intriguing that the update scheme performs better than the genetic algorithm when the percentage of broken memristors is between 50% and 80%, as shown in both the average number of failed trials (Fig. 3c) and the classification performance (Fig. 5). Here we present some tentative explanations. First, if we consider the perceptron's energy landscape, we can hypothesize that as more nodes get "stuck", the dimension of the configuration space reduces, making the landscape "simpler", therefore diminishing the advantage of the genetic algorithm in escaping the local minima. Secondly, since it does not use the local properties of the energy landscape, the genetic algorithm is not as efficient as the update method in optimizing the resistance values of non-damaged memristors. We can see from Fig. 4a that when the percentage of damaged memristors exceeds 50%, the fraction of non-damaged memristors set to the extreme values increases rapidly in the update scheme, while their fraction in the genetic algorithm scheme only rapidly increases after 80% of memristors are damaged. Notably, this range (50% to 80%) is also where we observe better performance from the update scheme. We may hypothesize that an effective way for a perceptron to accommodate the damaged nodes is by pushing the non-damaged nodes to their extreme values. Although more studies are needed, these two hypotheses nevertheless suggest that local schemes may be more beneficial for mitigating accumulated failures in hardware-based neuromorphic systems where the range of parameters in each node are physically constrained.

In summary, we implement the local update method and a genetic algorithm in a simulated memristor crossbar perceptron and compare their behaviors in image classification as the number of damaged memristors increases. We find that the update method is more resilient to the accumulation of failed memristors and exhibits an advantage in classification performance. Further investigations on the robustness of training algorithms in memristor crossbars against device failures are suggested.

## ACKNOWLEDGMENTS

The work was supported by the Faculty Research Grant from the College of Arts and Sciences, the University of Memphis.

## REFERENCES

[1] D. Ielmini and H.-S. P. Wong, Nat. Electron. **1**, 333 (2018).

[2] H. Yu, L. Ni, and H. Huang, in *Advances in Memristors, Memristive Devices and Systems*, edited by S. Vaidyanathan and C. Volos (Springer, New York, 2017), p. 275.

[3] M. A. Zidan, Y. Jeong, J. Lee, B. Chen, S. Hang, M. J. Kushner, and W. D. Lu, Nat. Electron. **1**, 411 (2018).

[4] K. Likharev, Sci. Adv. Mater. **3**, 322 (2011).

[5] M. Prezioso, F. Merrikh-Bayat, B. D. Hoskins, G. C. Adam, K. K. Likharev, and D. B. Strukov, Nature **521**, 61 (2015).

[6] C. Dias, J. Ventura, and P. Aguiar, in *Advances in Memristors, Memristive Devices and Systems*, edited by S. Vaidyanathan and C. Volos (Springer, New York, 2017), p. 305.

[7] S. H. Jo, T. Chang, I. Ebong, B. B. Bhadviya, P. Mazumder, and W. Lu, Nano Lett., **10**, 1297 (2010).

[8] S. Kumar, Z. Wang, X. Huang, N. Kumari, N. Davila, J. P. Strachan, D. Vine, A. L. D. Kilcoyne, Y. Nishi, and R. S. Williams, Appl. Phys. Lett. **110**, 103503 (2017).

[9] V. Ravi and S. R. S. Prabaharan, Far East J. Electron. Commun. **17**, 105 (2017).

[10] Y. L. Cun, J. S. Denker, and S. A. Solla, in *Advances in Neural Information Processing Systems 2*, edited by D. S. Touretzky, (Morgan-Kaufmann, San Francisco, 1990), p. 598.

[11] G. Taylor et al. in *Proceedings of the 33rd International Conference on International Conference on Machine Learning* Vol. 48 (JMLR.org, 2016) pp. 2722–2731.

[12] Y. A. LeCun, L. Bottou, G. B. Orr, and K. R. Müller, in *Neural Networks: Tricks of the Trade. Lecture Notes in Computer Science* Vol. 7700 (Springer, 2012), pp. 9-12.

[13] X. Liu and Z. Zeng, Complex Intell. Syst. (2021). https://doi.org/10.1007/s40747-021-00282-4.

[14] W. Schiffmann, M. Joost, and R. Werner, Optimization of the Backpropagation Algorithm for Training Multilayer Perceptrons (Technical Report, Institute of Physics, University of Koblenz, 1994)

[15] D. J. Montana and L. Davis, in *Proceedings of the 11th international joint conference on Artificial intelligence* Vol. 1 (Morgan Kaufmann, San Francisco, 1989), pp. 762–767.

[16] D. Whitley, Neural Networks, **1**, 230 (1988).

[17] P. G. Korning, Int. J. Neural. Syst. **6**, 299 (1995).

[18] P. Rocke, B. McGinley, F. Morgan, and J. Maher, in *Reconfigurable computing: Architectures, tools and applications* (Springer, 2007), pp. 373–378.

[19] G. Chakma, M. Adnan, A. R. Wyer, R. Weiss, C. D. Schuman, and G. S. Rose, IEEE J. Emerg. Sel., **8**, 125 (2018)

[20] F. Alibart, L. Gao, B. D. Hoskins, and D. B. Strukov, Nanotechnology **23**, 075201 (2012).

[21] F. Alibart, E. Zamanidoost, and D. Strukov, Nat. Commun. **4,** 2072 (2013).